\documentclass[aps,pre,preprint,epsfig,superscriptaddress,showpacs]{revtex4}
\usepackage{epsfig}
\usepackage{latexsym}
\usepackage{graphicx}
\usepackage{multirow}
\usepackage{threeparttable}

\usepackage{amssymb,amsfonts,amsmath}

\begin{document}
\title{Reweighted ensemble dynamics simulations: theory, improvement, and application}
\author{Linchen Gong}
\affiliation{Center for Advanced Study, Tsinghua University, Beijing 100084}
\author{Xin Zhou\footnote{Author to whom correspondence should be addressed; Electronic Mail: xzhou@ucas.ac.cn}}
\affiliation{School of Physics, University of Chinese Academy of Sciences, Beijing 100049}
\author{Zhong-Can Ou-Yang}
\affiliation{Center for Advanced Study, Tsinghua University, Beijing 100084}
\affiliation{Institute of Theoretical Physics, Chinese Academy of Sciences, Beijing 100086}

\date{\today}
 
\begin{abstract}
Based on multiple parallel short molecular dynamics simulation trajectories, we designed the reweighted ensemble dynamics (RED) method to more efficiently sample complex (biopolymer) systems, and to explore their hierarchical metastable states. Here we further present an improvement to depress statistical errors of the RED and we discuss a few keys in practical application of the RED, provides schemes on selection of basis functions, determination of free parameter in the RED. We illustrate the application of the improvements in two toy models and in the solvated alanine dipeptide. The results show the RED enable to capture the topology of multiple-state transition networks, to detect the diffusion-like dynamical behavior in entropy-dominated system, and to identify solvent effects in the solvated peptides. The illustrations serve as general applications of the RED in more complex biopolymer systems. 
\end{abstract}



\maketitle

\section{Introduction}
Exploring the high-dimensional conformational space of complex molecular systems has long been the focus of computer simulation studies. Two challenges exist under this topic, \emph{i.e.}, how to thoroughly sample the conformational space, and how to understand the equilibrium and dynamic properties based on lots of simulation data in hand. Although standard Monte-Carlo (MC) and molecular-dynamics (MD) simulation methods have been widely applied and achieved great successes, for many complex systems, such as glassy systems, phase-transition systems, and solvated biopolymers, sufficient sampling is by no means an easy task.  The rugged energy landscapes in the systems frequently traps simulation trajectories in a small conformational region, disapproving the thorough exploration of the whole conformational space within limited simulation time. Due to those, lots of various advanced simulation techniques, such as umbrella sampling~\cite{TorrieV1977,BartelsK1998,MaragakisVK2009}, J-walk~\cite{FrantzFD1990}, multicanonical ensembles~\cite{BergN1991}, simulating tempering~\cite{ST1, ST2}, conformational flooding~\cite{Grubmueller1995}, hyperdynamics~\cite{Voter1997b,MironF2004,ZhouJKZR2006}, parallel tempering~\cite{SugitaO2000,XuZO2012}, Wang-Landau sampling~\cite{WangL2001b} and its derivatives~\cite{YanP2003,KimSK2006}, meta-dynamics~\cite{LaioP2002}, multidimensional generalized-ensemble methods~\cite{MitsutakeO2009}, have been developed and applied to overcome the difficulties. New hardware~\cite{Shaw2010} and parallel computational technique~\cite{Wu2014} were also designed to extend MD simulations to reach microsecond scale in protein systems.  
In the other hand, due to the growing power of massively distributed computation, the ensemble dynamics (ED)~\cite{Voter1998,ShirtsP2001,HuangBGP2008}, 
which independently runs multiple simulations in distributed computers, provides an alternative way for abridging the gap between simulation power and practical interest. 

Besides the efficiently sampling, the proper analysis and understanding systems from simulation data are also not a trivial task. 
The major purpose in data analyse is to search for a coarse-grained description of the conformational space. In the traditional way, a few (usually only one or two) knowledge-based reaction coordinates (or order parameters) are first defined, then the equilibrium (and dynamic) properties of the system are achieved by mapping sampled configurations in the low-dimensional reaction-coordinate space. However, this method may suffer from the arbitrariness in the selection of reaction coordinates, and give an oversimplified description of the system~\cite{Noe2008}. Recently, a systematic metastable-state-network-based view brings prominent improvement~\cite{BeckerK1997,Wales2001,KrivovK2004,RaoC2004,GfellerRCR2007,NoeHSS2007,ChoderaSPDS2007,Prada-GraciaGEF2009,Bowman2013,WEBER2013}. Due to being theoretically transparent and inclusive of nearly all the useful information, the network model has received much attention. 
 
In a previous work~\cite{GongZ2009}, we have proposed a method, which was previously denoted as the weighted ensemble dynamics, now renamed as the reweighted ensemble dynamics (RED), to enhance sampling and to form metastable-state transition network~\cite{GongZ2010}. The RED is following the spirit of ED, which generated multiple short simulation trajectories from dispersively-chosen initial conformations, then applies a network-based analyse to get equilibrium and dynamics properties from all of these trajectories. In the present paper, we further improve and validate the RED, involving: (i) present a scheme to select basis functions and judge their completeness; (ii) discuss the unique free parameter in the RED; (iii) provide an improvement to depress statistical errors. We illustrate the improved RED in a few different systems, give the topology structure of multple-state transition network, predict the diffusion-like dynamical property in entropy-dominated systems. More importantly, we apply the RED to analyze effects of different aspects (degrees of freedom, interactions etc.) separately, such as the solvent effects in the solvated dipeptide. 
 
The article is organized as follows. In sect.~\ref{sec:theory}, we briefly represent the main formula of the RED, then we discuss a few keys in the RED, such as selection of basis functions, the effects of unique parameters, and the improvement of statistical errors. In sect.~\ref{sec:results}, we illustrate the application of the RED in two-dimensional multiple-well potentials and in solved dipeptides. A short conclusion is given in sect.~\ref{sec:conclud}. The simulation and analyse details are presented in the Appendix. 
 
\section{Theory and Method}
\label{sec:theory}
In the RED, multiple MD (or MC) trajectories, $\{q_i(t)\}, i=1, \cdots, p, t \in [0,\tau]$, are independently generated from initial conformations, denote as $\{q_{i}(0) = q_{i,0}\}, i=1, \cdots, p$, under the same simulation condition, respectively. Each trajectory involves $m$ conformations, $\{q_{i}(t=j \Delta) = q_{i,j}\}, j=0, \cdots, m-1$, thus total $m \times p$ conformations are sampled. Here $q$ is a simple notation of configurational coordinates. $\Delta$ is the time interval of samples, and $m \Delta = \tau$. We denote the distribution of samples in a single trajectory as $P_i(q) = \frac{1}{m} \sum_{j=1}^{m} \delta(q - q_{i,j})$, the initial distribution of all the trajectories as $P_{init}(q) = \lim_{p \rightarrow \infty} \frac{1}{p} \sum_{i=1}^{p} \delta(q - q_{i,0})$. 
  
\subsection{Theory of the RED}
\label{subsec:wed}
Usually, while $\tau$ is shorter than the equilibrium time of systems, a single trajectory does not reach the equilibrium, \emph{i.e.}, $P_i(q) \neq P_{eq}(q)$, but we can use all of these trajectories to reproduce the equilibrium distribution,  
\begin{equation}
\label{eq:aver_w} 
P_{eq}(q) = \frac{\sum_{i} w_{i} P_i(q) }{\sum_{i} w_{i}},  
\end{equation}
where $w_i$ is the weight of the $i$th trajectory, which is due to the deviation of the initial distribution from the equilibrium distribution. 
A possible selection of $\{w_i\}$ is,  
\begin{eqnarray}
w_i =  \frac{P_{eq}(q)}{P_{init}(q)}|_{q=q_{i,0}}. 
\end{eqnarray} 
Thus, we have a linear equation about $\{w_i\}$,  
\begin{eqnarray}
w_i = \frac{1}{p} \sum_j \Gamma_{ij} w_j, 
\label{eq:linearequation}
\end{eqnarray}
where $\Gamma_{ij} = \frac{P_{j}(q)}{P_{init}(q)}|_{q =q_{i,0}}$, and $\sum_j w_j = p$. 

Due to the finite size of samples $\{q_{i,j}\}$, Eq.~(\ref{eq:aver_w}) should be understood as, 
$\langle A(q) \rangle_{eq} = \frac{1}{p} \sum_{i} w_{i} \langle A(q) \rangle_i$, for any $A(q)$, while $\langle{}A(q)\rangle_i$ is estimated as $\frac{1}{m} \sum_{j=0}^{m-1} A(q_{i,j})$. Similarly, the ratio of two distributions is understood as, $\langle \frac{P_{j}(q)}{P_{init}(q)} A(q) \rangle_{init} = \langle A(q) \rangle_{j}$,  while $\langle \cdots \rangle_{\alpha}$ is estimated by average in the corresponding conformation samples. 
By expanding on a complete set of conformational functions (named as basis functions) $\{A^{\mu}(q)\}, \mu=1,2,\cdots$, we have, 
\begin{eqnarray}
{\Gamma}_{ij} = \sum_{\mu,\nu} g_{\mu\nu} A^{\mu}(q_{i,0}) \langle A^{\nu}(q) \rangle_{j}. 
\label{eq:matrix-element}
\end{eqnarray}
Here $g_{\mu\nu}$ is the inverse of $g^{\mu\nu} \equiv \langle A^{\mu}(q) A^{\nu}(q) \rangle_{init}$ which is estimated by the average in the initial sample. 
 
In practical application, instead of $A(q_{i,0})$ in estimate of $\Gamma_{ij}$, we use the average of $A(q)$ in a short $\tilde t$-length initial segment of trajectory (the first $m_0$ configurations with $m_0 << m$ or ${\tilde t} << \tau$, \emph{i.e.}, $A(q_{i,0}) \leftarrow \langle A(q) \rangle_{i^{+}} = \frac{1}{m_0} \sum_{j=1}^{m_0} A(q_{i,j})$, then 
\begin{eqnarray}
{\Gamma}_{ij} = \frac{1}{p} \sum_{\mu,\nu} \langle {\hat A}^{\mu}(q) \rangle_{i^{+}} \langle {\hat A}^{\nu}(q) \rangle_{j}. 
\label{eq:transform-matrix}
\end{eqnarray}
Here $\{{\hat A}^{\mu}(q)\}$ is the orthonormalized basis functions satisfied $\langle {\hat A}^{\mu}(q) {\hat A}_{\nu}(q) \rangle_{init+} = \delta_{\mu\nu}$ with $P_{init+}(q)=\frac{1}{p} \sum_i P_{i^{+}}(q)$. 
Thus, Eq.(\ref{eq:linearequation}) can be rewritten as, 
\begin{equation}
\label{eq:sym_lin_eq} H \vec{w}  = 0.
\end{equation}
Here $H \equiv G^{T} G$ is the variance-covariance matrix of the trajectory-mapped vectors 
\begin{eqnarray}
\vec{G}^{i} = (G_{i1},\cdots,G_{i,p})^{T}, 
\label{eq:trajectory-vector}
\end{eqnarray}
where $G_{ij} \equiv {\Gamma}_{ij} - \delta_{ij}$ with $i=1, \cdots, p$, and $\vec{w}=(w_1,\ldots, w_p)^T$. 
Eq.~(\ref{eq:sym_lin_eq}) is the main formula of the RED. 
The $\vec{w}$ corresponds to the ground state of matrix $H$ whose number of degenerated ground states gives the number of disconnected metastable states under the timescale of the individual trajectory $\tau$, the existence of a small fraction of transition trajectories among states can slightly split the ground state of $H$. More details are shown in our previous work~\cite{GongZ2009}. 

\subsection{Practical schemes of the RED}
\label{subsec:basis}
The selection of basis functions is a key in the practical application of RED. In our previous works~\cite{GongZ2009,GongZ2010}, we briefly discussed the 
selection of basis functions, and used some triangle functions as basis functions like Fourier expansion. Here we more generally discuss the key point of the RED. 
 
\subsubsection{Selection of basis functions}
Although the expansion in the RED required a complete set of basis functions in principle, it is usually not necessary to include too many basis functions in practice. 
Usually, it is easy to choose the initial configurations so that its distribution $P_{init+}(q)$ is already local equilibrium, thus only slow processes need to reweight by the RED. Therefore $\Omega(q)$ in the all-atomic $q$ space can be replaced by $\Omega(x) = \frac{P_{eq}(x)}{P_{init+}(x)}$ in a coarse-grained $x$ space, where $x$ represents the slow (coarse-grained) degrees of freedom, $P(x)$ is the corresponding distribution function in the $x$ space. For example, in proteins, if we throw away the first a few nanosecond relaxation simulation and reset the zero point of time, it is safe to suppose only some slow degrees of freedom, such as torsion angles of backbone, positions of side-chain atomic groups, be $x$, thus only functions in the $x$ space should be applied as basis functions. In many cases, systems might be reach equilibrium within the $\tau$ time (length of individual trajectory) except only a few collective variables, thus we even apply the RED in a low-dimensional space.   
In addition, we usually suppose there are many potential-energy basins (or super-basins) separated by high (free) energy barriers in the conformational space, each (super-)basin corresponds a metastable state wherein a $\tau$-length trajectory is sufficient to reach local equilibrium. Thus, $\Omega(x)$ looks like a step-like function, which is almost constant in each metastable state, varies only happening at boundaries between states. 
We approximately have 
\begin{eqnarray}
\Omega(x) \approx \sum_{\alpha} \Theta_{\alpha}(x) \omega_{\alpha},
\label{eq:step-weight}
\end{eqnarray}
where $\Theta_{\alpha}(x)$ is character function of the $\alpha$ metastable state, which is unity if $x$ is inside the state, zero otherwise.  
The set of character functions of states $\{\Theta_{\alpha}(x)\}$ provides a natural set of basis functions to expand $\Omega(x)$, thus the required number of basis functions is equal to the number of states. Since the boundaries of these states are unknown a priori, we usually use far more basis functions to linearly combine the step-like $\omega(x)$ function. 
We may use some usual collective variables with clear physical meaning as basis functions. In biological macromolecules, dihedral angles and their analytical functions, pair distances between some key atomic groups, potential energy or its parts, solved area of proteins, radius of gyration, native contact number, number of hydrogen, etc., can be applied as basis functions. The variance-covariance matrix $g^{\mu\nu}$ will ensure the consistent consideration of different classes of basis functions, and provide the measure of their relative importance by the orthonormalization process. In addition, the number of independent basis functions is also limited by the finite size of $init+$, thus more selected basis functions will be discarded after considering the matrix $g_{\mu\nu}$, then do not significantly contribute to final results. 
Another possible scheme about the selection of basis functions is to use local functions in the conformational space. For example, for a set of sampled configurations, $\{x_i\}, i=1,\cdots, M$, ones may group them into some subsets by clustering algorithms based on their geometric similarity, and use the character functions of clusters as basis functions. Thus the number of basis functions is the number of clusters, $n_{cl} =\frac{M}{m}$. Here $M$ and $m$ are the total number of sampled conformations and the (average) number of conformations in each cluster. The idea is actually already applied in MSM to divide the whole conformational space into metastable states~\cite{NoeHSS2007,ChoderaSPDS2007,Prada-GraciaGEF2009,Noe2008,Bowman2013,WEBER2013}. In the paper, we use physical-meaningful collective variables as basis functions, the application of the sample-based discrete cluster functions will appear elsewhere. 
 
\subsubsection{Completeness of basis functions}
The completeness of selected basis functions can be directly test based on sampled configurations in the RED. For a set of basis functions $\{A^{\mu}(x)\}$, 
we can estimate the difference between two samples (or the corresponding distributions), such as the initial distribution $P_{init+}(x)$ and the equilibrium distribution $P_{eq}(x)$, 
\begin{equation}
S_{init+,eq}^2 = \sum_{\mu,\nu} g_{\mu\nu} \langle {A}^{\mu}(x) \rangle_{eq} \langle {A}^{\nu}(x) \rangle_{eq}. 
\end{equation}
The definition gives a measurement about the similarity of samples, can be generally applied to compare two samples. For example, we had used the formula to optimise coarse-grained models of the all-atomic water~\cite{LvZ2015}. 
 For assessing the completeness of basis functions in expansion of $\frac{P_{eq}(x)}{P_{init+}(x)}$, a reasonable method is to check the convergence of 
 $S_{init+,eq}^2$ while adding more and more basis functions. 
 
To calculate $S_{init+,eq}^2$, the sample for equilibrium distribution $P_{eq}(x)$, thus the trajectory weights $\{w_i\}$, are necessary. We propose two methods to bypass this difficulty. First, before reweighting, $P_{eq}(x)$ is replaced by ${\bar P}(x) = \frac{1}{p} \sum P_i(x)$, and we roughly select basis functions. Second, after reweighting, $S_{init+,eq}^2$ can be calculated with the resulting trajectory weights. This procedure can be used to test the former selection. If the rigorously calculated $S_{init+,eq}^2$ has reached its saturation region with previously roughly selected basis functions, the results should be satisfiable. However, this posterior examination is restricted to the cases that the trajectory weights can be uniquely determined. In practice, the distribution ${\bar P}(x)$, has the same local equilibrium characteristics in each metastable region as $P_{init+}$ and $P_{eq}$, thus $\Omega_{init+,{\bar P}} = \frac{ {\bar P}(x)}{P_{init+}(x)}$ is similarly a step function, which is also constant in each meta-stable region, only the constant values are different from those of $\Omega_{init+,eq}$. Thus we believe the a priori test is already enough, and the following results are all calculated by this way. 
 
Since the statistical errors exist due to the finite sample sizes, $S^2$ would not strictly stop growing with the number of basis functions, it just increases slowly compared to the initial growing phase with increasing basis functions. Similarly, when large error occurs, $S^2$ may abnormally increase within the saturation region. 
We design a scheme to get rid of the statistically unreliable basis functions. We orthonormalize the functions $\{A^{\mu}(x)\}$, leading to the explicitly identifiable value and error contribution from the orthonormalized basis functions to $S^2$. In current paper, we implement the Gram-Schmidt orthonormalization method to one-by-one derive out the set of orthonormalized basis functions $\{ {\hat A}^{\mu}(x)\}$, thus $S^2=\sum_{\mu} s_{\mu}^2$, with $s_{\mu}^2 = \langle {\hat A}^{\mu}(x) \rangle_{eq}^2$. To estimate the error of $S^2$, we approximately estimate the error of $s_{\mu}$, and combine them by
\begin{equation}
Err(S^2)=\sqrt{\sum_{\mu} 4 s^{2}_{\mu} Err^{2}(s_{\mu})}.
\end{equation}
Here, $Err()$ denotes the statistical error of the quantity in brackets. For each orthonormalized basis function, $\hat{A}^{\mu}$, if $Err(s_{\mu})$ is larger than $|s_{\mu}|$, \emph{i.e.}, the relative error of $s_{\mu}$ is larger than $1.0$, this basis function will not be selected in expansion. One-by-one selection will give out the final basis function set and the estimated error of the completeness measure $S^2$.

The scheme introduced here provides us the knowledge about how many basis functions are enough in the RED. After orthonormalization, the contribution of the basis functions to $S^2$ clearly reflect their relative importance in trajectory reweighting. The basis functions that contribute more are more important for portraying the difference between $P_{init+}(x)$ and $P_{eq}(x)$, and the ones that contribute quite small can be linearly expanded by other existing basis functions. 
Furthermore, various kinds of basis functions can be safely added to investigate the related phenomena. For illustration, we will show that the basis functions of dihedral angles and those of distances between heavy atoms can be consistently treated in calculation, and the effect of solvent on the conformation of biological macromolecules can be studied by additional basis functions characterising the solute-solvent relation. 

\subsubsection{The effects of free parameter $\tilde{t}$}
\label{sec:tildet}
There is only one free parameter in the RED, ${\tilde t}^{*}=\tilde{t}/\tau$, for collecting the $init+$ samples in Eq.(\ref{eq:transform-matrix}). Large ${\tilde t}^{*}$ may bring systematical errors, and small ${\tilde t}^{*}$ will reduce the sample number in $init+$, leading to a larger statistical uncertainty. $\tilde{t}^{*}=0.01 \sim 0.1$ is usually applied in the RED with satisfiable results obtained. We also point out that, for the purpose of discovering the states in conformational space, the results are not sensitive to ${\tilde t}^{*}$. 

$\tilde{t}$ actually reflects the timescale of the process that we are interested in. If it were set to the single trajectory length $\tau$, then the trajectory weights solved by Eq.~(\ref{eq:sym_lin_eq}) will be equal to each other. In this case, we are focusing on the processes with a timescale much longer than $\tau$, thus all the trajectories are still in the relaxation to local equilibrium, and should have the same weight. Usually, if a system can be divided into several metastable states under the focused timescale, $\tilde{t}$ should be selected comparable to or smaller than the smallest life time of these states. Given that the kinetic states of the system are frequently unknown, $\tilde{t}$ should be selected relatively short. In practice, it is possible to use shorter and shorter $\tilde{t}$ and check the convergence of the obtained free energy surface. Thus, a proper $\tilde{t}$ may be found as a compromise between correctness and statistical reliability (small fluctuation of trajectory weights). Another practical consideration is that, a few percent of $\tau$ would be the safe choice for $\tilde{t}$. This is because the transition events with timescales prominently smaller than $\tau$ have reached equilibrium in single trajectory (thus do not need reweighting), we only need to focus on the ones with time-scales similar to or slightly smaller to $\tau$. We usually choose $\tilde{t}/\tau$ in the interval of $[0.01, 0.1]$, and the results are satisfiable. In the paper, we will explicitly show that the $\tilde{t}^{*}$ within the range is not sensitive to our results. 
 
\subsection{Improvement of the RED} 
\label{subsec:statistics}
For decreasing statistical errors in the RED, we can generate a few trajectories from each initial configuration, and specify the same weight to the trajectories from the same initial configuration. Suppose there are $n_s$ initial configurations (denoted as $q_{10},q_{20},\ldots,q_{n_s0}$ in the following), and $m_1,m_2,...,m_{n_s}$ trajectories initiated respectively from these initial configurations (the total number of trajectories is still $p$), we have
\begin{equation}
\label{eq:sym_mult_eq}
\sum_{k=1}^{n_{s}} (\sum_{j} G_{ij} {\tilde T}_{jk}) w_{k} = 0
\end{equation}
where $w_{k}$ is the weight for all the $m_{k}$ trajectories initiated from $q_{k0}$, ${\tilde T}$ is a $p \times n_s$ matrix with zero and unity elements. $\tilde{T}_{jk}$ is unity only if the $j$th trajectory starts from the $k$th initial configuration. Since the number of equations is more than the number of trajectory weights, we estimate Eq.~(\ref{eq:sym_mult_eq}) by minimizing $I= | G \tilde{T} \vec{w} |^{2}$. The vector of trajectory weights, $\vec{w}$, corresponds to the ground state of matrix $H \equiv {\tilde T}^{T} G^{T} G {\tilde T}$. For the newly defined $H$ matrix, the number of the degenerated ground states still corresponds to the number of disconnected metastable states under the timescale of the individual trajectory. The existence of a small fraction of transition trajectories can slightly split the ground state of $H$. We name this method as SI1. 

Alternatively, we can merge the trajectories with the same initial configuration to a single one, and perform the standard RED analysis. Concretely speaking, we can calculate the average of certain variable in the initial-segment sample set, $init+$, by,
\begin{eqnarray}
 \langle A \rangle_{init+} = \frac{1}{n_{s}} \sum_{i=1}^{n_{s}} \langle A \rangle_{q_{i0}^{+}},
\end{eqnarray}
where $\langle A \rangle_{q_{i0}^{+}}$ means the average of $A$ over the initial segments of all the trajectories with the initial configuration $q_{i0}$. By doing so, more samples are included in calculation, and effectively, both the initial region and the whole area in the conformational space which the trajectory explores become statistically more reliable. This method is named as SI2. We will illustrate the application of SI1 and SI2 for depressing the large statistical noise in the entropy-dominated systems where the differences between trajectories within a metastable state might be in comparison with that in different states. 

\section{Results}
\label{sec:results}

\subsection{Toy models}

\begin{figure*}
\includegraphics[width=5in]{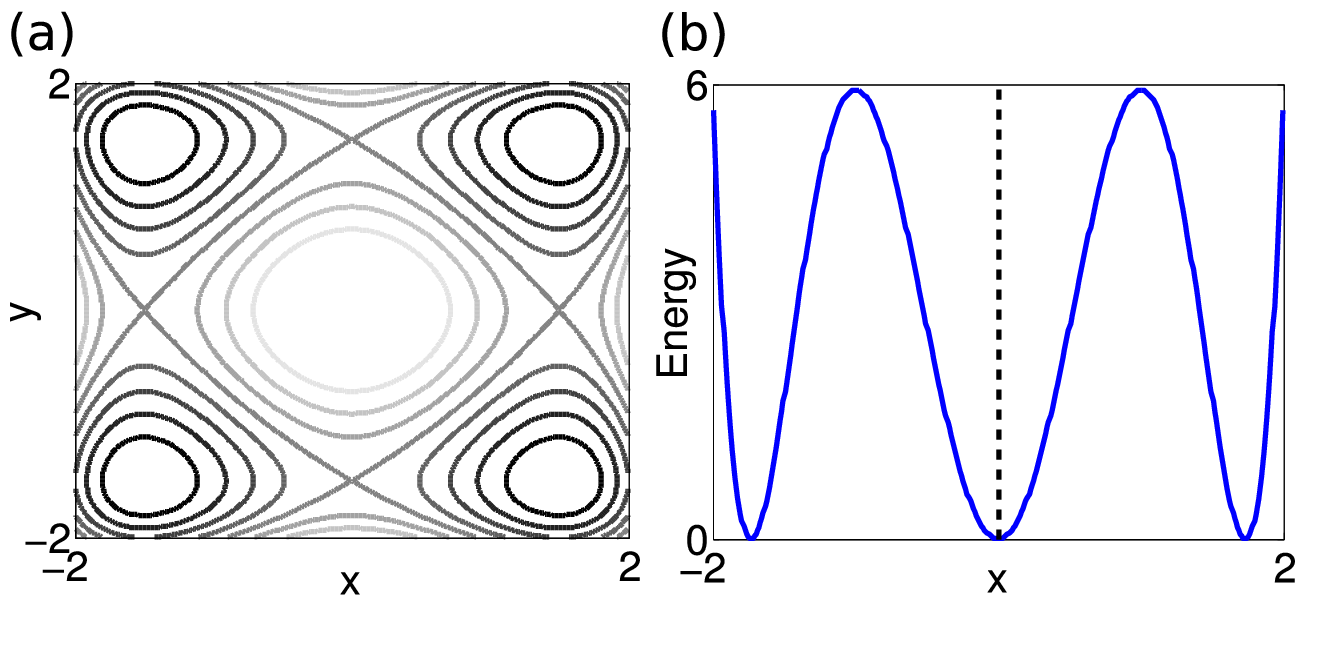}
\caption{\  \ 
( Color online) The 2D potentials. (a) The TDQW potential shown as contour map. The darker regions correspond to potential wells, and the lighter regions correspond to potential barriers. (b) The MHAT potential projected along one-dimension. Rotating around the dashed line reproduces the 2D potential.
\label{fig:2d_pot}
}
\end{figure*}

We first illustrate the RED in two simple toy models. The first one is a two-dimension (2D) system with four same potential wells, denoted as Two-Dimension-Quad-Well (TDQW), see Fig.~\ref{fig:2d_pot}(a). The second model is a 2D system with a Mexican-Hat potential (MHAT) which has two states, see Fig.~\ref{fig:2d_pot}(b). The outer state is wide (larger entropy) and the relaxation time inside the state is comparable to the transition time between the state and the inner state. Simulation details and the applied basis functions are shown in Appendix. 

\subsubsection{Multiple-state transition network} 
\begin{figure*}
\includegraphics[width=6in]{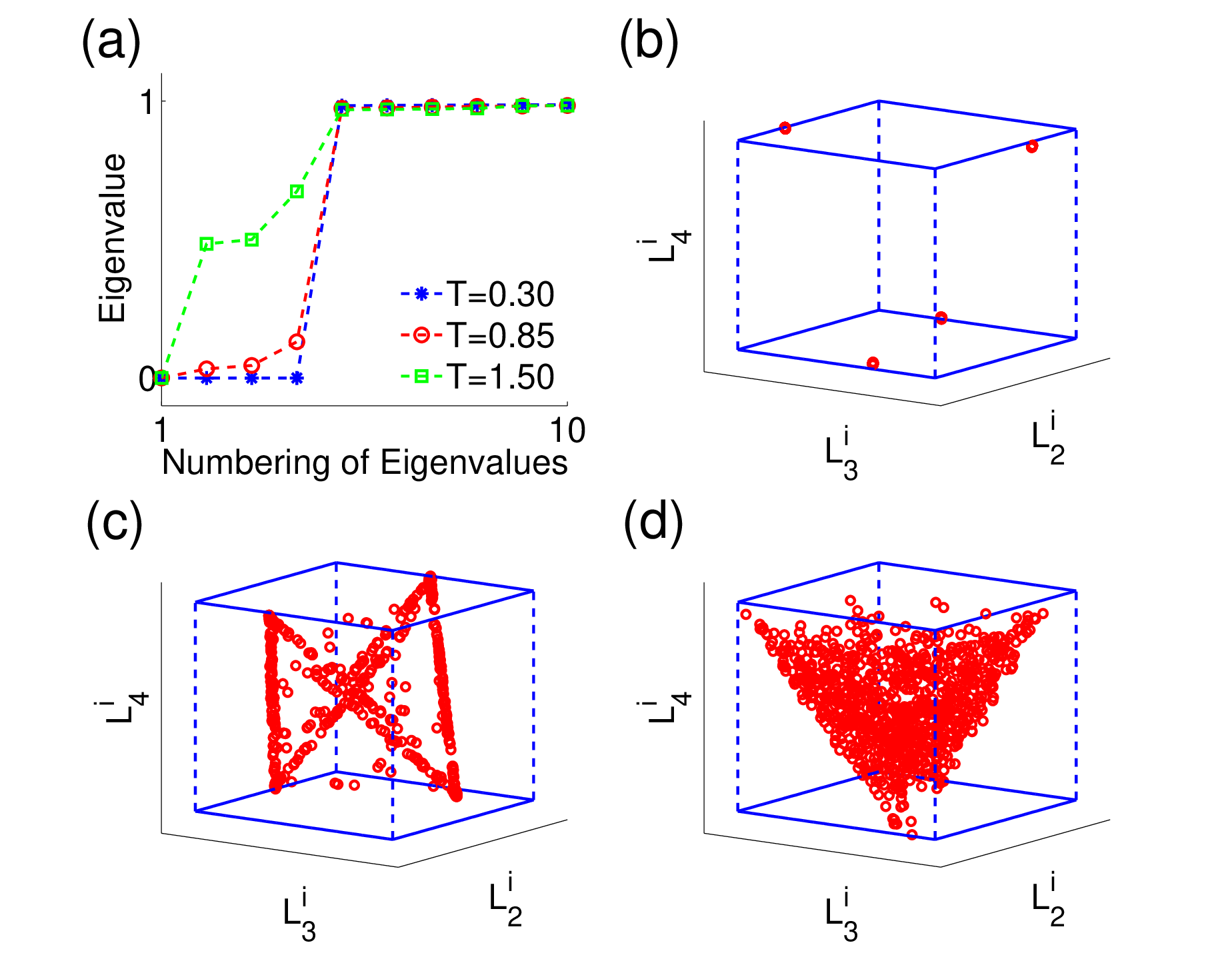}
\caption{\  \  
(Color online) The eigenvalues and projection maps for TDQW potential under different temperatures. (a) The first ten eigenvalues of $H$ matrix calculated at different temperatures. For the three temperatures of $0.3$(b), $0.85$(c) and $1.50$(d), the projection of $\{\vec{G}^i\}$ to the second, third and fourth eigenvectors of $H$ are also plotted. 
 \label{fig:vecval_mdmw}
}
\end{figure*}

In TDQW system, the eigenvalues of the matrix $H$ in Eq.(\ref{eq:sym_lin_eq}) are plotted in Fig.~\ref{fig:vecval_mdmw}(a). As expected, there are four zero eigenvalues at temperature $0.3$, implying the four disconnected states. At higher temperatures, the second to fourth eigenvalues deviate from zero to larger values due to the increasing fraction of the interstate transition trajectories. More illustrative perspective can be obtained from Fig.~\ref{fig:vecval_mdmw}(b), (c), and (d), where the trajectories are mapped into the three-dimensional space by projecting the trajectory-mapped vectors, $\{\vec{G}^i\}$ defined in Eq.(\ref{eq:trajectory-vector}), to the second, third and fourth eigenvectors of $H$, denoted as $L^i_{\alpha}\equiv\vec{G}^i\cdot\vec{u}_{\alpha}$, where $\vec{u}_{\alpha}$ is the $\alpha$th eigenvector of $H$. At temperature $0.3$, the trajectories agglomerate into four disconnected clusters that locate respectively at the four vertexes of a tetrahedron. Inside each cluster are the trajectories isolated in the same potential well. At temperature $0.85$, part of the trajectories can climb over the potential barriers to the other states. As a result, the points corresponding to the transition trajectories are found to gather along the straight lines that shoot out of the vertexes, and the ones corresponding to the non-transition trajectories are still locating on the vertexes. 
In Fig.~\ref{fig:vecval_mdmw}(c), each vertex only connects to two neighboring vertexes, correctly reflecting the topology of TDQW system. At temperature $1.5$, it is relatively easier for the trajectories to transfer between different states. The tetrahedron structure is still preserved, and its interior is filled by multiple-transition trajectories which transition among more than two states. 
 
\subsubsection{Entropy-dominated states}
\begin{figure*}
\includegraphics[width=6in]{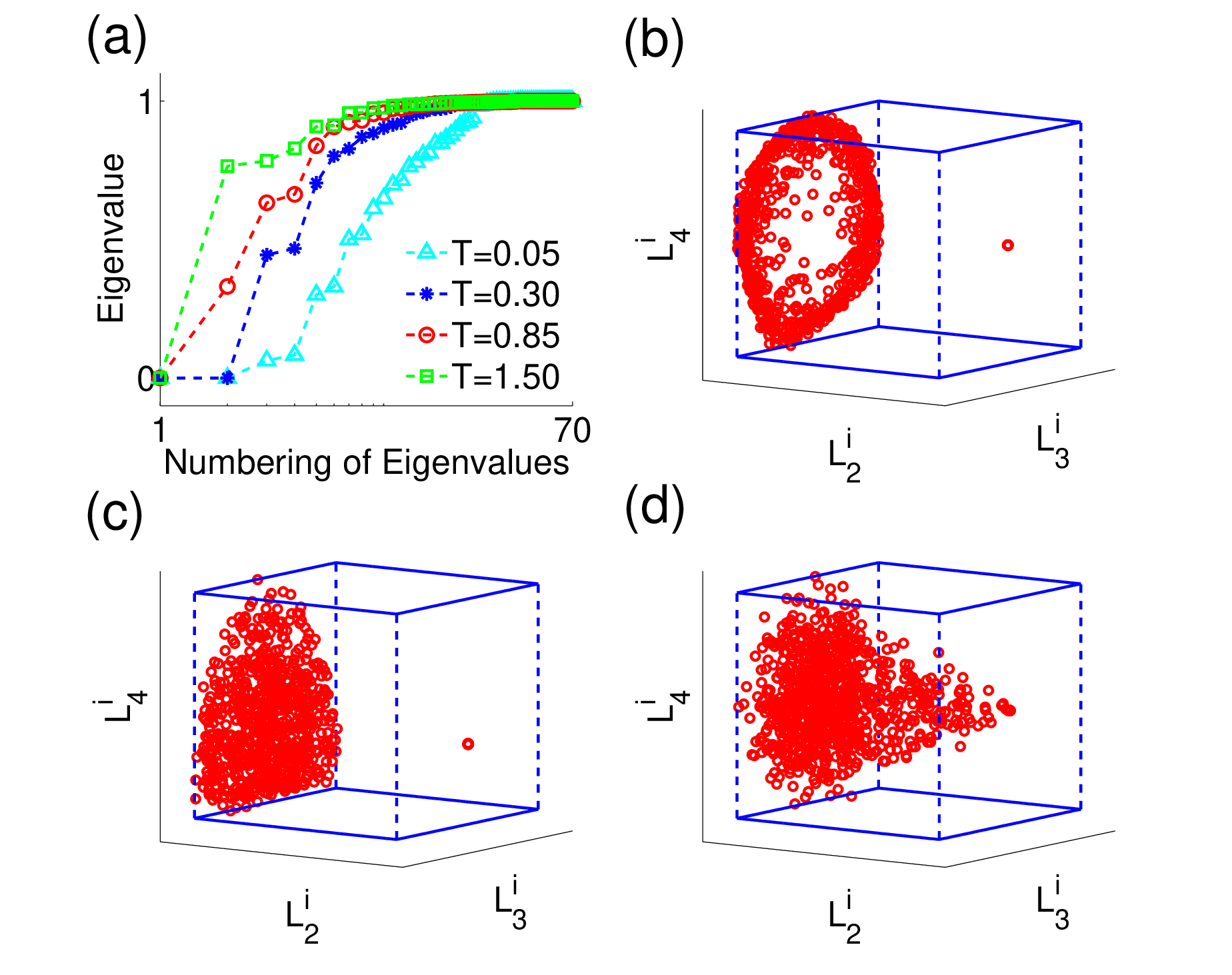}
\caption{ \  
(Color online) The eigenvalues and projection maps for MHAT potential under different temperatures. (a) The first seventy eigenvalues of $H$ matrix calculated at different temperatures. For the three temperatures of $0.05$(b), $0.30$(c) and $0.85$(d), the projection of $\{\vec{G}^i\}$ to the second, third and fourth eigenvectors of $H$ are also plotted. 
 \label{fig:vecval_mhat}
}
\end{figure*}

In MHAT system, the $70$ smallest eigenvalues of $H$ are shown in Fig.~\ref{fig:vecval_mhat}(a). Different from TDQW, the eigenvalues of $H$ gradually increase from zero at the low temperature $T=0.05$, and lots of small eigenvalues exist. The number of the eigenvalues that obviously deviate from $1.0$ are always more than the number of physical states in MHAT system, till the temperature is high enough that there is only one zero-eigenvalue apparently different from $1.0$. This behavior simply reflects the entropic nature of the outer potential well. At low temperature, a single trajectory can not reach local equilibrium in the state, consequently a few small eigenvalues exist, like there are many sub-states in the well. At higher temperature, the trajectories that start from the outer potential well usually climb over the potential barrier before reaching local equilibrium within the state, thus the inter-trajectory difference will not be extinguished, until the two states in MHAT kinetically become one state at high enough temperature. The projections of $\{G^i\}$ onto the eigenvectors of $H$ are shown in Fig.~\ref{fig:vecval_mdmw}(b), (c), and (d). At both temperatures $0.05$ and $0.30$, the trajectories are divided into two groups. In one group, the trajectories dispersively locate on the same plane which is perpendicular to the $L^i_2$ axis. These trajectories are found to be isolated in the outer potential well. In the other group, the trajectories concentrate to nearly one point. They are found to be isolated in the inner potential well. These two groups of trajectories are clearly differentiated by the projections along the second eigenvector of $H$, \emph{i.e.}, the one indicating the most important contribution to ergodicity broken. At $T=0.05$, each trajectory only explores very limited region in the outer state, leading to the distribution on the perimeter of a circle. At $T=0.30$, since each trajectory can explore larger area, the circle has been filled with points. At $T=0.85$, the trajectories are found to locate in a cone. The points in the bottom surface and the vertex of the cone correspond to the non-transition trajectories inside the outer and inner potential wells, respectively, and the other points are transition trajectories between the two states. Once again, the trajectory-projected space clearly provides the topological of this system. 

\subsubsection{discussions} 
\begin{figure*}
\includegraphics[width=6in]{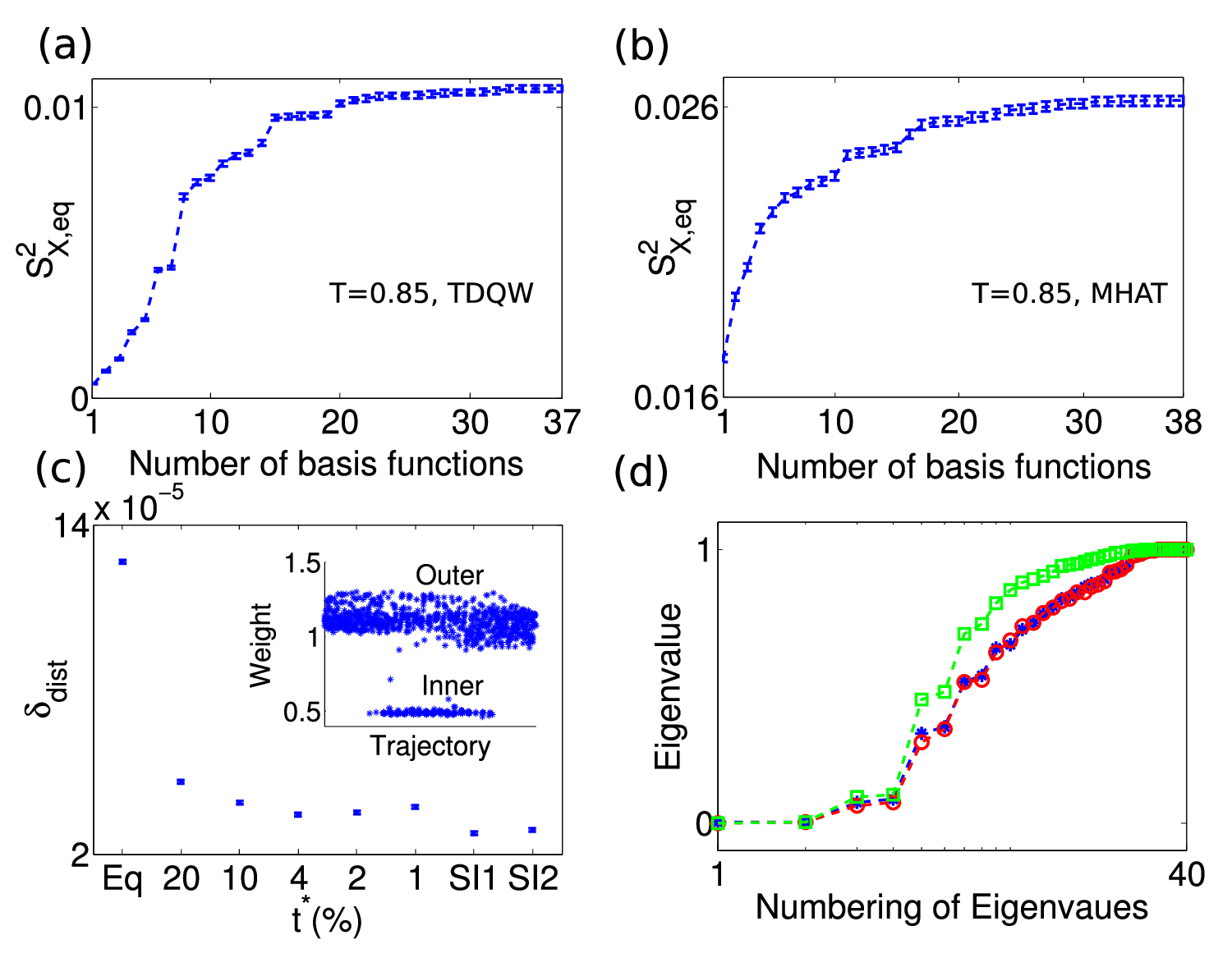}
\caption{ \  
(Color online) The $S^2$ curve, effect of $\tilde{t}$ and illustration of statistics improvement. The behavior of $S^2$ with increasing number of basis functions are shown in (a) (TDQW) and (b) (MHAT). (c) The difference between the reconstructed equilibrium distribution and the theoretical distribution are shown for MHAT potential at $T=0.85$. The horizontal axis denotes the method for reproducing the equilibrium distribution. `Eq' means the results without reweighting (\emph{i.e.}, $w_i =1$),  numbers mean the selected $\tilde{t}^*=\tilde{t}/\tau$. `SI1' and `SI2' label the two methods for statistics improvement, respectively. The trajectory weights calculated with $\tilde{t}^*$ equal to $0.01$ are plotted as inset. In figure (a), (b) and (c), the statistical errors are also shown as error bars. (d) The first $40$ eigenvalues of $H$ for MHAT potential at $T=0.05$. The ones calculated with normal data set (stars), multiple trajectory data set (one-fifth initial configurations and five trajectories for each initial configuration) with the standard RED (circles), multiple trajectory data set with the SI2 (squares) are shown.
 \label{fig:s2_tt_is}
}
\end{figure*}

In the two 2D models, triangle functions in the 2D space are applied as basis functions, details are presented in Appendix. 
We did the $S^2$ analysis mentioned in sec.\ref{subsec:basis} at $T=0.85$. For both of the two systems, $S^2$ initially grows fast with increasing number of basis functions, then the growth slows down and shows a saturation behavior of $S^2$ as expected [see Figs.~\ref{fig:s2_tt_is}(a) and (b)]. Once $S^2$ has approximately reached the platform region, more basis functions will not change the calculation results any more (data not shown). Thus, for TDQW and MHAT systems, about $24$ basis functions are sufficient.

We also study the effect of free parameter $\tilde{t}$ in the RED with the simulation data of MHAT at $T=0.85$. In the case, since there is only one zero eigenvalue of $H$, the whole free energy surface can be reconstructed. We uniformly divide the whole conformational space into small cells, $k=1, \ldots, N_{p}$ ($N_{p}$ is selected as $1600$ here, {\it i.e.}, $40$ bins along each dimension), and the distribution probability in each cell is reconstructed with different $\tilde{t}^* = \tilde{t}/\tau$. We define the function $\delta_{dist}(\tilde{t}^*)$ as 
\begin{equation}
\delta_{dist}(\tilde{t}^*)=\sqrt{\frac{1}{N_p}\sum_{k=1}^{N_p}(P_{k}(\tilde{t}^*)-P_{k}^{theo})^2}
\end{equation}
to measure the difference between the reconstructed distribution and the theoretical distribution.
Here $P_{k}(\tilde{t}^*)$ is the distribution in the $k$th cell which is reproduced by the RED using
$\tilde{t}^*$. $P_{k}^{theo}$ is the theoretical distribution in the cell. As shown in Fig.~\ref{fig:s2_tt_is}(c), $\delta_{dist}(\tilde{t}^*)$ indeed decreases with shorter $\tilde{t}^*$ in calculation. For $\tilde{t}^*$ in the $[0.01,0.1]$ interval, the $\delta_{dist}(\tilde{t}^*)$ values are almost same, and are prominently smaller then the value of the non-weighted distribution of the simulation data. In the non-weighted (all $w_{i}$ is unity) distribution, the major deviation from theoretical distribution resides in the inner potential well of MHAT, where the distribution is too high for the non-weighted data. In the reweighting process, the trajectories that start from the inner potential well are usually specified lower weights [see Fig.~\ref{fig:s2_tt_is}(c), inset], thus the over occupation in the inner state can be offset. The other variation in trajectory weights further adjust the intrastate distribution for both the inner and the outer states.

\subsubsection{Improvement of statistical errors} 
We test the statistics improvement methods, SI1 and SI2. In here and below, we always randomly select one-fifth of the initial configurations, then five trajectories are spawned from each configurations to test the improvement. At $T=0.85$, we apply both SI1 and SI2 to reconstruct the energy surface of MHAT. From Fig.~\ref{fig:s2_tt_is}(c), it can be seen that both methods correctly reproduce the energy surface with even smaller deviation from the theoretical one, as compared to the standard RED. At $T=0.05$, the eigenvalues of the $H$ matrix are calculated with SI2 method. As shown in Fig.~\ref{fig:s2_tt_is}(d), no matter one trajectory or multiple trajectories are generated from one initial configuration, the eigenvalues calculated by the standard RED are almost same. However, analyzing by the SI2 method significantly alters the structure of eigenvalues. By generating more trajectories from single initial configuration, and combining these trajectories together in analysis, the statistic of single trajectory is prominently improved, and the difference between trajectories within the outer state are reduced to large extent. As a result, some of the eigenvalues increases, in particular for the ones deviating but close to $1.0$. In other words, the SI2 method has depressed the statistical noise in the spectral analysis of $H$, leaving the physically relevant states recognized.

\subsection{Solvated alanine dipeptide} 
\begin{figure*}
\includegraphics[width=5in]{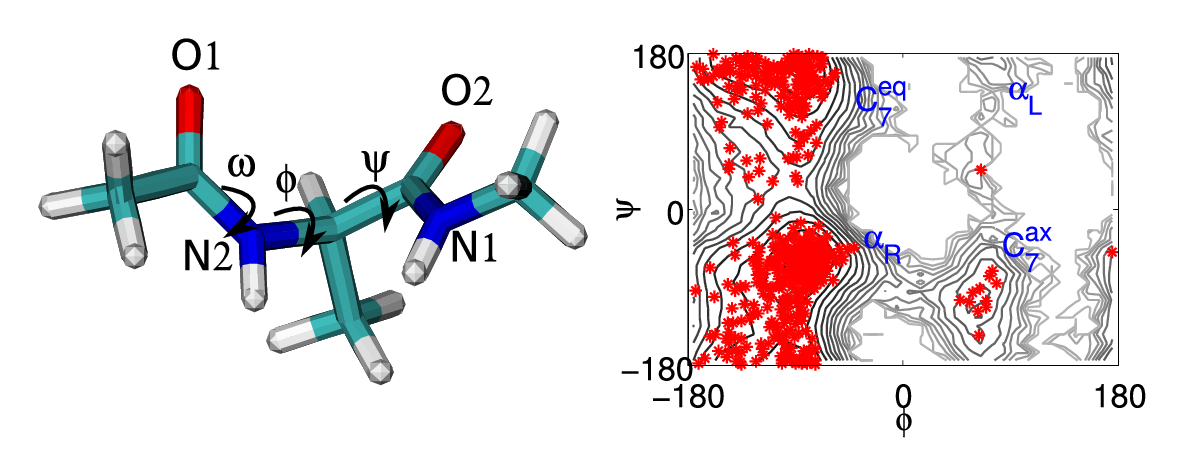}
\caption{\  \  
(Color online) Illustration of alanine dipeptide molecule and the initial conformations of the RED simulation. The alanine dipeptide is shown in the left panel, with some dihedral angles and atoms labeled. The initial conformations projected to the $\phi$-$\psi$ plane are shown in the right panel. The background is the free energy surface at $T=600$ ${\rm K}$, with the free energy wells labeled. 
 \label{fig:alad_desc}
}
\end{figure*}

The simulated system together with the initial configurations of simulation are shown in Fig.~\ref{fig:alad_desc}, where various notations of the atoms, dihedral angles and free energy wells of this molecule are introduced. More details of simulations are presented in Appendix. 
We use two different classes of basis functions to focus on the effects of different physical variables on the observed metastable states. 
 
\subsubsection{Dihedral angles}
With thorough investigation, the two dihedral angles $\phi$ and $\psi$ are the major coordinates to characterise the conformational motion of alanine dipeptide molecule. Same as that in the previous study~\cite{GongZ2009}, we adopted the first-to-fourth order two-dimensional trigonometrical functions of $\phi$ and $\psi$ as the basis functions. The $10$ smallest eigenvalues of $H$ are shown in Fig.~\ref{fig:bond_angle}(a). The smallest zero eigenvalue is intrinsic due to the construction of $H$. The second eigenvalue corresponds to the kinetic separation between the region with $\phi>0$ (containing $C^{ax}_7$ and $\alpha_L$) and the region with $\phi <0$ (containing $C^{eq}_7$ and $\alpha_R$). Consequently, by projecting the trajectory-mapped vectors $\{\vec{G}^i\}$ to the second eigenvector of $H$, \emph{i.e.}, through $\{L^i_2\}$, the trajectories can be clearly classified into the above mentioned two regions. The third eigenvalue of $H$ corresponds to the partially kinetic connection between the $C^{eq}_7$ and $\alpha_R$ states. Only a fraction of the trajectories in $C^{eq}_7$ can climb over the free energy barriers to the $\alpha_R$ state, and \emph{vice versa}. Thus, we may use the $\{L^{3}_i\}$ to classify the trajectories that are confined in the super-state involving $C^{eq}_{7}$ and $\alpha_{R}$ into three groups. Since the two potential wells are well defined with their internal relaxation time smaller than the transition timescale, we can manipulate the trajectories to predict the transition times and locate the transition state ensemble~\cite{GongZ2009}. The fourth smallest eigenvalue of $H$ is also smaller than $1.0$. Although it no longer indicates new state in the conformational space and should be taken as the reflection of statistical difference between trajectories, we recently find that, by observing $\{L^4_i\}$, two trajectories that transiently walk into the $\alpha_L$ region can be picked out. One of the two trajectories initially located in the $\alpha_L$ region, then quickly moved into $C^{ax}_7$, another just occasionally visited $\alpha_L$. Actually, at the temperature of $300$ $\rm{K}$, the $\alpha_L$ region seems not very stable with the applied force field. The other eigenvalues of $H$ are very close to $1.0$ without observable meaningful information. 
 
\subsubsection{Distances between heavy atoms}
\label{sec:inatomdist}
\begin{figure*}
\includegraphics[width=6in]{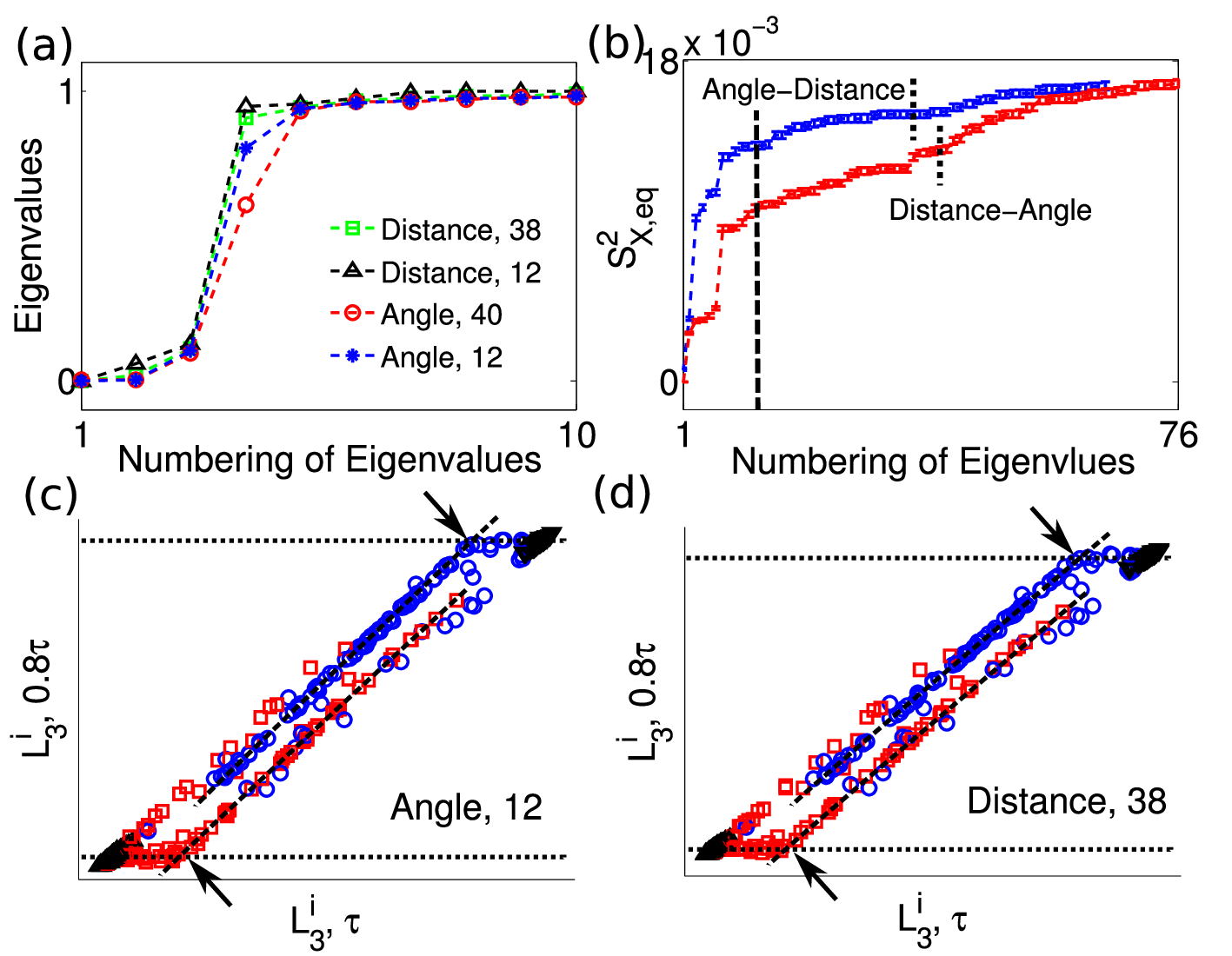}
\caption{\  \   
(Color online) RED analysis with the basis functions of dihedral angles and inter-atomic distances. Shown are the results for solvated alanine dipeptide, which is simulated at $T=300 {\mathrm K}$ with the modified force field. (a) The eigenvalues of $H$ matrix calculated with different sets of basis functions, including the full set of inter-atomic distances (squares), $12$ inter-atomic distances (upward triangles), first-to-fourth order trigonometrical functions of  $\phi$ and $\psi$ (circles) and first-to-second order trigonometrical functions of $\phi$ and $\psi$ (stars). (b) The  behavior of $S^2$ is shown either with the inter-atomic distances(Distance-Angle) at first, or with the trigonometrical functions of $\phi$ and $\psi$ at first (Angle-Distance). The $\{L^3_i\}$ values calculated with truncated trajectories ($80$ percent length) versus the ones calculated with full trajectories are shown. The results are obtained either with the first-to-second order trigonometrical functions of $\phi$ and $\psi$ (c), or with the inter-atomic distances (d). The arrows label the intersection points between the dashed and dotted lines.
 \label{fig:bond_angle}
}
\end{figure*}

We also test the effects of distances between atoms in the backbone of dipeptide by using them as basis functions. We compare two aspects, \emph{i.e.}, the eigenvalues of $H$ and the manipulations on $\{L^3_i\}$ for extracting the kinetic information between $C^{eq}_7$ and $\alpha_R$. There are $10$ non-hydrogen heavy atoms in alanine dipeptide. Except for the directly bonded atoms, $36$ pairs of atoms exist, corresponding $36$ interatomic distances. Together with the two distances for the possible intramolecular hydrogen bonds (see Table~\ref{tab:basis}), we select the $38$ distances as basis functions. The resulting eigenvalues of $H$ are shown in Fig.~\ref{fig:bond_angle}(a). While the three smallest eigenvalues are almost same with the results obtained with the functions of dihedral angles, the fourth eigenvalue is prominently lifted to a value closer to $1.0$. However, the discrimination of the two $\alpha_L$-visited trajectories is still possible. We plot the $\{L^i_3\}$ values calculated with the trajectories truncated to $80$-percent length ({\it i.e.}, the $0.2\tau$ ending segment of the trajectories are discarded in analysis) versus the ones calculated with full trajectories~\cite{GongZ2009} in Fig.~\ref{fig:bond_angle}(d). 
For comparison, the same figure generated with the first-to-second order two-dimensional trigonometrical functions of torsion angles are shown in Fig.~\ref{fig:bond_angle}(c). The two figures have quite similar structure. The points corresponding to the non-transition trajectories in $\alpha_R$ and $C^{eq}_7$ are plotted as upward and downward (black) triangles, respectively. The points corresponding to the transition trajectories that start from $\alpha_R$ and $C^{eq}_7$ are plotted as (red) squares and (blue) circles, respectively. 
The points along the inclined dashed lines correspond to the trajectories with only one time transition, and the points along the horizontal dotted lines correspond to the trajectories that do not happen transition within the initial $0.8\tau$. Thus the intersection between the dotted and dashed lines correspond to a single-transition trajectory with its transition happened exactly at $0.8\tau$. These points can be used to predict the times the transitions happen in the single-transition trajectories, consequently the transition state ensemble can be constructed. 
All the other points correspond to the trajectories with even number of transitions which all happen within $0.8 \tau$, or with early transitions occurred within $\tilde{t}$, or the multiple transition trajectories with transition happened both within $0.8 \tau$ and after $0.8 \tau$. More details about detecting transition time thus transition conformation ensemble from the truncated-trajectory plots are presented in Ref.~\cite{GongZ2009}. The similarity between Fig.~\ref{fig:bond_angle}(c) and Fig.~\ref{fig:bond_angle}(d) suggests both the torsion angles and the inter-atomic distances can be applied as basis functions.

\subsubsection{Solvent molecules}
For solvated biological macromolecules, the conformational change is sometimes correlated to the surrounding solvent molecules. In the RED, since the solvent-related properties can be selected as basis functions, it becomes possible to systematically one-by-one check and filter out the important collective variables. 
We study the dipeptide system with $91$ basis functions to search for the solvent-related effects. These basis functions are classified into three groups, which respectively reflect the internal freedoms of alanine dipeptide (denoted as P), the solute-solvent relation (denoted as PW) and the structure of bulk solvent (denoted as W). The details of these basis functions are listed in Table~\ref{tab:basis}. The eigenvalues of $H$ are shown in Fig.~\ref{fig:temp_eigval}(a). As can be seen, at $T=300 \mathrm{K}$, the inclusion of the solvent-related basis functions does not affect the spectral property of $H$. Particularly, the eigenvalues that are prominently smaller than $1.0$ do not change, reflecting the unchanged physical division of the conformational space. Only a few eigenvalues pretty close to $1.0$ are slightly lowered due to the added basis functions. The above results suggest that, for the current system, the solvent related properties have almost reached equilibrium within the timescale of single trajectory. Consequently, these properties contribute quite a little to the inter-trajectory difference, and are unlikely to be strongly coupled with the dynamic process of the solute molecule under the simulated timescale. 

\subsubsection{Low temperature}

We also apply the RED to study the low-temperature ($T=150 \mathrm{K}$) properties of the dipeptide. 
We note that, at this temperature, the physical phenomena are quite complicated; the force field and the simulation procedure that we adopt may not reflects the physical reality. Here, we are only meant to illustrate the application of the RED in the extreme condition, instead of providing physical insight into the system.

\begin{figure*}
\includegraphics[width=6in]{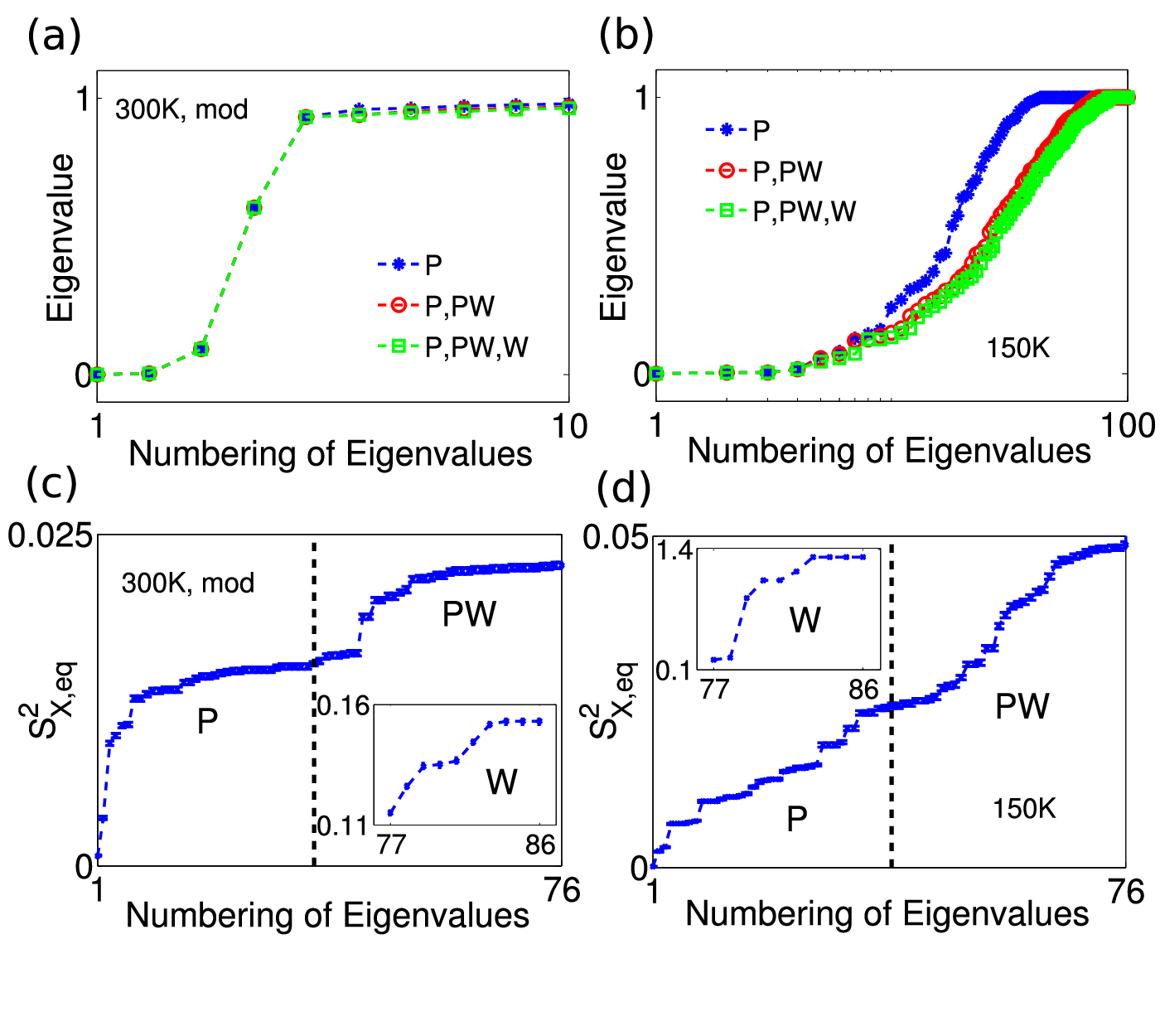}
\caption{\  \   
(Color online) The spectral properties of $H$ calculated with the basis functions including solute-solvent  relation. Shown are the results for the system of solvated alanine dipeptide. The eigenvalues of $H$ calculated with  different groups of basis functions are shown in (a) ($300$ ${\rm K}$ simulation with modified force field) and (b) ($150$ ${\rm K}$ simulation with standard forcefield). The functions used in calculation are labeled in the legend. The symbols `P', `PW', and `W' mean the three sets of basis functions included in calculation, the internal degrees of freedom of dipeptide, the solvent-solute interaction, and the bulk water, see the main text for more details. The behavior of $S^2$ with increasing number of basis functions are shown in (c) ($300 {\mathrm K}$ with modified force field) and (d) ($150 {\mathrm K}$ with standard force field). The insets of (c) and (d) show the part of $S^2$ curve that is related to the functions of solvents.  
 \label{fig:temp_eigval}
}
\end{figure*}

The eigenvalues of the $H$ matrix are calculated with the $91$ basis functions listed in Table~\ref{tab:basis}, and are plotted in Fig.~\ref{fig:temp_eigval}(b). 
The solvent-related quantities do affect the eigenvalues of $H$ now, leading to even more small eigenvalues. Correspondingly, the $S^2$, as shown in Fig.~\ref{fig:temp_eigval}(d), almost always increases with the number of basis functions. At such a low temperature, the free energy surface becomes quite rough, the diffusive dynamics~\cite{Zwanzig1988} in the conformational space slows down seriously. As a result, the simulation trajectories may only explore a small region in the conformational space. Some degrees of freedom, like the solvent motion around the solute molecule and the conformational motion perpendicular to the $\phi$-$\psi$ plane, are no longer equilibrated in single trajectory. Thus, the inter-trajectory difference becomes larger, more states are detected automatically, and more basis functions are required to expand the rugged distribution of the samples, leading to the abnormal behavior of the eigenvalues of $H$ and $S^2$.

\begin{figure*}
\includegraphics[width=6in]{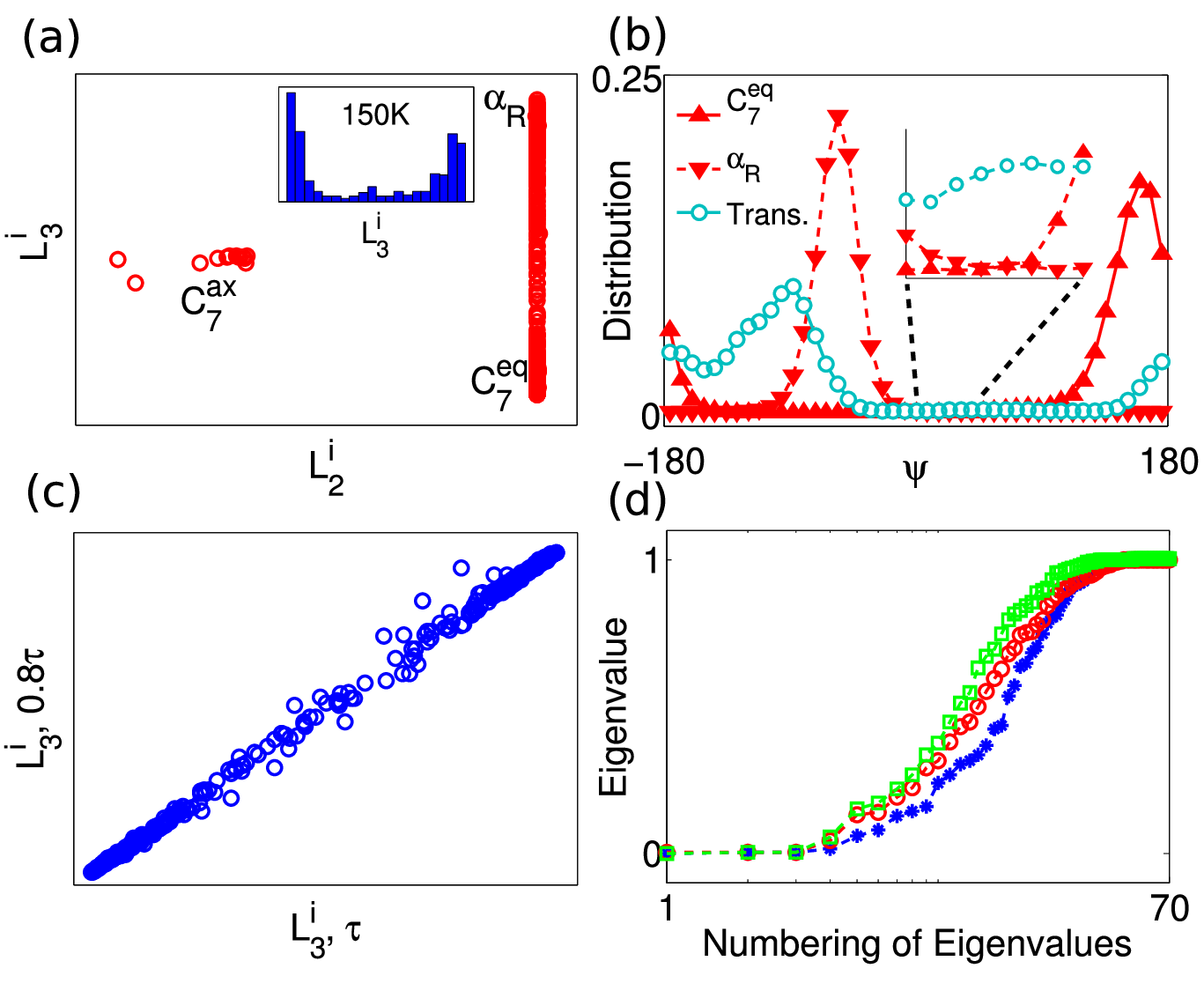}
\caption{\  \   
(Color online) Results for the low-temperature simulation of alanine dipeptide. (a) The projection of  trajectories into the space of $L^i_2$ and $L^i_3$. The inset is the histogram of $\{L^i_3\}$ values. The different  states, into which the trajectories are roughly classified, are labeled. (b) The distributions along the $\psi$ angle in the groups are shown in (a). Shown are the distributions for the trajectories in the leftmost two bins (downward triangles), the right most two bins (upward triangles) and the other trajectories (circles). (c) The $\{L^3_i\}$ values calculated with truncated trajectories ($80$ percent length) versus the ones calculated with full trajectories. (d) The first $70$ eigenvalues of $H$ calculated with normal data set (blue stars), multiple trajectory data set (one-fifth initial configurations and five trajectories for each initial configuration) with the standard RED (red circles), multiple trajectory data set with SI2 method (green squares) are shown. 
 \label{fig:low_temp}
}
\end{figure*}

Manipulating the eigenvectors of $H$ can provide further information of the simulation data. The projections of $\{\vec{G}^i\}$ to the second and the third eigenvectors of $H$ pick out two completely separated states, $C^{ax}_{7}$ and the super-state containing $\alpha_{R}$ and $C^{eq}_{7}$ [Fig.~\ref{fig:low_temp}(a), at the current low temperature, the $\phi$ and $\psi$ angles may not be enough to characterize the conformational motion, thus the notations, $C^{ax}_{7}$, $\alpha_{R}$ and $C^{eq}_{7}$, are used mainly for the purpose of illustration]. By the aid of the histogram of $\{L^i_3\}$ [see Fig.~\ref{fig:low_temp}(a), inset], we can roughly classify the trajectories inside the super-state into three groups, \emph{i.e.}, the trajectories in the leftmost two bins, in the rightmost two bins and the remaining ones. The conformational distribution of trajectories in different groups are shown in Fig.~\ref{fig:low_temp}(c). The two groups of the leftmost or rightmost bins both possess single peak distribution, either in $C^{eq}_{7}$ region or $\alpha_{R}$ region, just like the non-transition trajectories identified in the simulations at higher temperature. In contrast, the trajectories in the remaining group have a significant occupation in the intermediate region between $C^{eq}_{7}$ and $\alpha_{R}$. Obviously, these trajectories are trapped at the intermediate region due to slow dynamics, and are different from the transition trajectories identified with the similar procedure at $300 {\mathrm K}$. We plot the $\{L^i_3\}$ calculated with $80$ percent trajectory length with the ones with full trajectory length in Fig.~\ref{fig:low_temp}(c). The behaviors presented in Fig.~\ref{fig:bond_angle} no longer exist. In terms of kinetics, the transition between the two regions, $\alpha_R$ and $C^{eq}_7$, can neither be approximated as Markovian process nor described with the first-order transition process any more. Consequently, with the current simulation data, it is only possible to reconstruct the intrastate equilibrium distribution in the local regions of conformational space, or to study the more localized kinetic behavior.

Due to the numerous local minimums at the low temperature, the $\alpha_R$ and $C^{eq}_7$ regions are just like two entropic-dominated states. We also test the statistics improvement method in this system. With one-fifth initial conformations and five trajectories from each initial configuration, the inter-trajectory difference, thus the statistical noise, is indeed suppressed as shown in Fig.~\ref{fig:low_temp}(d).

As  a summary, in the dipeptide system at $T=300 {\mathrm K}$ with modified potential, we find the redundant basis functions (including functions of both the dihedral angles and the inter-atomic distances) are consistently treated in the RED, and the inclusion of the solvent-related basis functions gives out the expected results. At pretty low temperature of $150 {\mathrm K}$, the roughness of the free energy surface is reflected in the results of RED. The conformational space can still be roughly
divided into parts, however, the kinetics becomes too complicated to be unambiguously characterized. 

\section{Conclusion}
\label{sec:conclud}
 As a systematic method for exploring the conformational space, the RED does not require much priori knowledge and approximation, thus its results can honestly reflect the thermodynamic and kinetic information of the system. For biological macromolecules, the usually expected hierarchical structure can be extracted out
step-by-step. For these complex systems, the structure of the conformational space may be much more complicated, these  examples studied here become quite relevant for physically interpreting the results. Along with the applications, we design the $S^2$ quantity for checking the completeness of the selected basis functions, and the schemes for statistics improvement. These tactics are also proved to be helpful for portraying the conformational space.

\acknowledgements{ 
The work is supported by NSFC under Grant No. 11175250. X.Z thanks the financial supports of the Hundred of Talents Program in Chinese Academy of Sciences. 
L.C. and X.Z. also thank the support of Max Planck Society and the Korea Ministry of Education, Science and Technology, part of the work was done while they worked at Asia Pacific Center for Theoretical Physics, Korea under the support.}

\appendix 
\section{Toy models} 
\subsection{Simulation details}
For the 2D toy models, TDQW and MHAT, simulations are restricted in the square region of $[-2,2]\times[-2,2]$. Reflecting boundary condition is imposed. The over-dampened Langevin equations
\begin{eqnarray}
\frac{dx}{dt}=-\frac{1}{\gamma}\frac{\partial{}U(x,y)}{\partial{}x}+\sqrt{\frac{2k_BT}{\gamma}}\xi_1(t)\nonumber\\
\frac{dy}{dt}=-\frac{1}{\gamma}\frac{\partial{}U(x,y)}{\partial{}y}+\sqrt{\frac{2k_BT}{\gamma}}\xi_2(t)\nonumber\\
\end{eqnarray}
is adopted to generate the dynamical trajectories. Here $\gamma$ is the frictional coefficient. 
$\xi_1(t)$ and $\xi_2(t)$ are the white noises satisfying 
$\langle\xi(t)_i\xi(t')_j\rangle=\delta(t-t')\delta_{ij}$ with $\langle\rangle$ denoting the
ensemble average of noise. We simply take the Boltzmann constant $k_B$ and $\gamma$ as unity to get the dimension-reduced units for time, position and temperature.

The potential of TDQW is
\begin{widetext}
\begin{displaymath}
U(x,y)=\left\{\begin{array}{lllll}
              \infty & , & |x|>2 & \text{or} & |y|>2 \\
              5(\frac{4}{9}x^2-1)^2+5(\frac{4}{9}y^2-1)^2 & , & |x|\leq2 &
              \text{and} & |y|\leq2\\
              \end{array}
\right..
\end{displaymath}
\end{widetext}
The potential of MHAT is
\begin{equation}
U(r)=40\cdot(\frac{1}{27}r^6-\frac{2}{9}r^4+\frac{1}{3}r^2),\quad{}r^2=x^2+y^2\nonumber
\end{equation}

For both systems, at each temperature, $900$ initial configurations are randomly selected, then the RED simulations are performed, with each trajectory simulated for $50$ dimensionless time units. The coordinates are recorded every $0.01$ time units. The $\tilde{t}/\tau$ is selected as $0.04$, if not further mentioned.
In TDQW system, three temperatures of $0.30$, $0.85$ and $1.50$ are discussed.  
For MHAT system, the RED simulations are performed at four temperatures of $0.05$, $0.30$, $0.85$, and $1.50$.
 
\subsection{Basis functions} 
We apply the following two-dimensional triangle functions
\begin{eqnarray}
\label{eq:hd_base}
&\sin{}[(m+n)x],\cos[(m+n)x],&\nonumber\\
&\sin{}[(m+n)y],\cos[(m+n)y],&\nonumber\\
&m+n>0&\nonumber\\
&\sin{}(mx)\sin{}(ny),\sin{}(mx)\cos{}(ny),&\nonumber\\
&\cos{}(mx)\sin{}(ny),\cos{}(mx)\cos{}(ny),&\nonumber\\
&m\geq{}1,n\geq{}1&
\end{eqnarray}
as basis functions. Here, $(x,y)$ are the 2D Cartesian coordinates multiplied by a transformation factor $\frac{\pi}{2}$, $m$ and $n$ are non-negative integers. We define the summation of $m$ and $n$ in Eq.~(\ref{eq:hd_base}) as the order of these functions, and adopt the first-to-fourth order functions (totally $40$ basis functions) in our analysis. 

\section{Solvated alanine dipeptide}
For the system of explicitly solvated alanine dipeptide, we generated $500$ MD trajectories at $T= 300 \mathrm{K}$, each with $\tau=600 \mathrm{ps}$ length in the NAMD package\cite{NAMD2005} with CHARMM27 force field. In the simulations, a few potential barriers were added on the dihedral energy terms of the standard force field to kinetically separate the conformational space at the room temperature. The detailed simulation settings is same as that in our pervious work~\cite{GongZ2009}. 
For the low temperature simulations, first the $500$ configurations selected from a $600 {\mathrm K}$ simulation in the NAMD package are first relaxed by simulating at $150 {\mathrm K}$ for $200 {\mathrm ps}$, then continuously simulate $\tau = 800 \mathrm{ps}$, and collect configurations every $0.2$ ${\rm ps}$ for subsequent analysis.

\begin{table*}
\centering
\begin{threeparttable}
\caption{\label{tab:basis} Basis Function List}
\begin{tabular}{p{3.5cm}p{2cm}p{10cm}}
\hline
Group & Number & Function \\
\hline
\multirow{2}{3.5cm}{Internal freedom\\ of  alanine dipeptide}
 & 40 & Two-dimensional trigonometric functions of $\phi$ and $\psi$. \\
 & 8  & $\rm{N}1-\rm{O}1$, $\rm{N}2-\rm{O}2$, $\rm{N}1\rm{H}-\rm{O}1$, $\rm{N}2\rm{H}-\rm{O}2$ and their squares*.\\
\hline
\multirow{5}{3.5cm}{Interaction between\\ solvent \\ and solute}
 & 2  & Interaction** between peptide and water. \\
 & 8  & Interaction between $\rm{O}1$, $\rm{O}2$, $\rm{N}1$, $\rm{N}2$ and water. \\
 & 2  & Water number around peptide ($3.3\AA$, $5.5\AA$).\\
 & 1  & Solvent accessible surface area.\\
 & 20 & Water number around $\rm{O}1$, $\rm{O}2$, $\rm{N}1$, $\rm{N}2$ ($3.3\AA$, $3.7\AA$, $4.1\AA$, $4.5\AA$, $4.9\AA$). \\
\hline
Solvent structure
 & 10 & Integration of the $\rm{OO}$ radial distribution function, $g(r)$, of the bulk water***, {\it i.e.}, $\int_0^{r_0}r^2g(r)dr$ ($r^0$ is selected to be $2.9\AA$, $3.3\AA$, $3.7\AA$, $4.1\AA$, $4.5\AA$, $4.9\AA$, $5.3\AA$, $5.7\AA$, $6.1\AA$ and $6.5\AA$). \\
\hline
\end{tabular}
 \begin{tablenotes}
  \item[*] The four atoms $\rm{O}1$, $\rm{O}2$, $\rm{N}1$ and $\rm{N}2$ are labelled in Fig.~\ref{fig:alad_desc}, left panel. $\rm{N}1\rm{H}$($\rm{N}2\rm{H}$) is the hydrogen atom bonded to $\rm{N}1$($\rm{N}2$). $\rm{O}1$-$\rm{N}1$ and $\rm{O}2$-$\rm{N}2$ may form intra-molecule hydrogen bonds, or may be connected by water bridge~\cite{Pappu2004}.
  \item[**] In here and below, interaction means the electrostatic energy part and the Van de Waals energy part.
  \item[***] These functions reflect the packing of bulk water.
 \end{tablenotes}
\end{threeparttable}
\end{table*}


\end{document}